\documentstyle[eqsecnum,aps]{revtex}
\begin{document}
\draft

\title{Dislocation mediated melting
near isostructural critical points}

\author{T. Chou and David R. Nelson}

\address{Department of Physics, Harvard University,
Cambridge, MA 02138}

\date{\today}

\maketitle

\begin{abstract}
We study the interplay between an isostructural critical
point and dislocation mediated two-dimensional
melting, using a combination of Landau and continuum
elasticity theory. If dislocations are excluded, coupling to
the elastic degrees of freedom leads to  mean field critical
exponents.  When dislocations are allowed, modified
elastic constants lead to a new phase buried in the solid
state near the critical point.  Consistent with a
proposal of Bladon and Frenkel,
we find an intervening hexatic phase near the critical
point of the first-order isostructural transition line.
Very close to the critical point, a
transition from the hexatic phase to a modulated bond
angle state is possible.  Ising transitions in
symmetrically confined two dimensional colloidal
crystals with attractive wall potentials require a different
model free energy, which we also discuss.
\end{abstract}

\section{Introduction}

First order phase transitions preserving
symmetry and possessing a critical point have
been well understood for many years, the most
common example being the discontinuous
liquid-vapor phase transition.  Such systems
can be brought continuously from one phase to
another by taking a thermodynamic path which
passes around the critical point along the
equation of state surface without sudden
changes in density. Near the critical point,
liquid-vapor systems show critical opalescence
among other properties peculiar to second
order phase transitions.

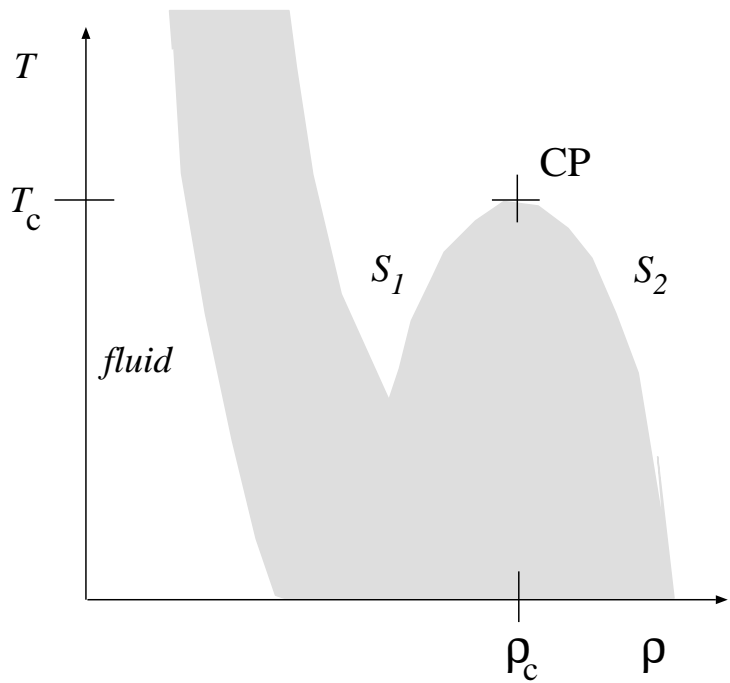
\begin{figure}
\caption{$T-\rho$ phase diagram of a system exhibiting
a solid-solid phase coexistence. (CP) is a critical point,
$S_{1}$ and $S_{2}$ are the low and high density solids
respectively.}\label{SSPHASE}
\end{figure}

Recently, attention has been focussed on {\it
solid-solid} phase transitions
\cite{FRENKEL0,FRENKEL1,TEJERO}.  Especially
interesting are isostructural transitions in
which the lattice spacing jumps
discontinuously, but the symmetry of the
lattice remains unchanged. A schematic phase
diagram as a function of temperature, $T$, and
particle density, $\rho$, is shown in Figure
\ref{SSPHASE}.  As in the liquid-vapor
transition, isostructural transitions can
have a critical point above which the
distinction between the isostructural phases
$S_{1}$ and $S_{2}$ vanishes.
Because the solid-solid coexistence region
is on the higher density (solid) side of the
fluid-solid coexistence region, a
fluid-solid-solid triple point becomes
possible.

Experimentally, bulk isostructural transitions
have been observed in high pressure studies of
Zr.  Above the fcc$\rightarrow \omega$ (at 6.7
GPa) and $\omega \rightarrow$ bcc (at 33GPa)
transitions, a lattice symmetry preserving
transition between two bcc phases occurs at
53GPa \cite{FCCBCC}. Another experimentally
convenient system with which to study
solid-solid transitions is colloidal
particles.  Colloids are often observed to be
in an ordered fcc or bcc lattice, useful for example, in
colloidal array filters \cite{CAF6}. The
interaction potential between colloid
particles can be chemically tailored to
control the range of interactions.  For
example, one can vary the length of grafted
polymers, or change solvent properties
\cite{COLLOIDS6}. Single layers of colloids
can be confined to study the transitions of 2D
solids. This has been achieved by squeezing
colloidal particles between two flat plates
\cite{CAM6} or floating them at an air-water
interface \cite{CHAN}.

The theoretical study of the statistical mechanics of
colloidal particles has utilized various analytical
\cite{TEJERO,ERNST,KIRKWOOD,TAREYEVA} and numerical
\cite{ALDER,FRENKEL3} methods. It is well established that
hard spheres, with or without a superimposed attractive
potential, can undergo a solid-fluid phase transition.
When the  attractive potentials are short-ranged enough,
experiments and theory demonstrate that the liquid-gas
coexistence vanishes \cite{FRENKEL3,GAST}. The
short-ranged attractive potentials required can be
achieved with colloidal particles or in computer
simulations.  A recent numerical study \cite{FRENKEL1}
using particles with very
short-ranged attractive potentials (attractive region
$<7\%$ of the hard core diameter), and at high densities,
show solid-solid phase
coexistence.  This behavior is consistent with
the variational treatment of Tejero {\it et.  al.}\cite{TEJERO}

In this paper, we explore how ideas of defect mediated
melting \cite{MELTREV} may be incorporated into the
schematic phase diagram of  particles with short-ranged
attractions. There is some evidence for a two-stage
dislocation mediated melting process in liquid crystals
\cite{LXTAL}, colloidal dispersions \cite{CAM6}, and
magnetic bubble domains \cite{WESTERVELT}, however, a
first order transition from solid to liquid (for example by
generation of grain boundaries) is also possible
\cite{MORF}. The study of defects away from an inherent
solid-fluid transition, deep in the solid phase, near a
solid-solid critical point, is a new opportunity to test
predictions of dislocation melting theory \cite{MELTREV}.

As pointed out by Bladon and Frenkel \cite{FRENKEL1}, an
intervening hexatic phase (the product of the first stage
in the two-stage melting scenario) is likely near a two-dimensional
isostructural critical point.  Chen {\it et.  al.}, studying an
isothermal, isobaric Lennard-Jones system, point out the
large system sizes required in searching for a metastable
hexatic phase \cite{CHEN}.  Nevertheless, simulation
measurements of elastic constants by Bladon and Frenkel
\cite{FRENKEL1} show that a dislocation instability
(presumably leading to a hexatic phase) {\it must}
develop  in the vicinity of a solid-solid critical point.
Since the compressibility diverges near the critical point
in Figure \ref{SSPHASE}, the effective bulk modulus must
soften \cite{FRENKEL1}.  This softening of the bulk
modulus promotes dislocation unbinding which signals
the solid-hexatic transition.  In this paper, we explore
these ideas using a simple physical model and determine
how the existence of a solid-solid critical point may
induce a nearby hexatic phase.  The effects of elastic
degrees of freedom on the Ising critical properties in the
{\it absence} of dislocations are also considered.

Finally, even though most simulations
exhibiting a solid-solid phase
transition rely on a very short-ranged
attractive potential, longer-ranged,
soft potentials may also show similar
behavior. For example, the
three-dimensional solid-solid transition
in dense Cs and Ce \cite{JAY} may be
pressure induced \cite{FRENKEL0,ALDER2}.
Two-dimensional realizations such as
confined colloids \cite{CAM6} and
colloids \cite{CHAN} or C$_{60}$
particles \cite{C60} at air-liquid
interfaces also possess complicated
short and long-ranged interactions.
However, the {\it two-dimensional}
melting outlined relies only on a
solid-solid critical point and  is not
sensitive to the microscopic mechanism
of particle interactions.

In the next section, we introduce two free energy
functionals including both elastic degrees of freedom and
the Ising-like order parameter describing the
isostructural solid-solid phase transition. Elastic
deformations and the Ising order parameter are coupled
via an interaction energy. We also briefly discuss
symmetrically confined colloidal crystals in 2D with
attractive wall potentials. If the wall potential leads to an
Ising-like symmetry-breaking, such that the crystal
prefers to be near one wall or the other, a new type of
isostructural critical point becomes possible.

Section III discusses the simplest model in the absence of
defects. Above the critical point, the Ising degrees of
freedom renormalize the elastic coefficients yielding an
effective bulk modulus.  Thermal fluctuations in the
(single-valued) strains renormalize the critical
temperature of the Ising degrees of freedom. We show
that the critical temperature renormalization is different
for the uniform and finite wavevector  modes in a manner
such that the Ising critical point acquires {\it mean-field}
behavior. The uniform, zero wavevector strains induce
an infinite-ranged interaction that insures
{\it classical} critical exponents in this
two-dimensional system.

In Section IV, the strains and Ising order parameter in the
presence of a single dislocation are studied by solving the
extremal (Euler) equations. The strain field configurations
are asymptotically identical to those of an uncoupled
model except they are functions of effective elastic coefficients. The
related Ising order parameter near the dislocation is also
determined. An energy-entropy argument for an isolated
dislocation leads to the melting criterion simulated by
Bladon and Frenkel \cite{FRENKEL1}.  Qualitative phase
diagrams containing solid and hexatic phases are derived
based on this criterion.

The free energy of the full interacting model is expressed
in terms of a collection of {\it many} dislocations in
Section V.  Renormalization group flows for the the
parameters in the presence of a few dislocations are also
derived.  Dislocation-renormalized elastic moduli yield a
melting criterion agreeing with simple energy-entropy
considerations.  The critical exponents of the
crystal-hexatic transition are the same as for
conventional dislocation-mediated melting.
\cite{DRN79,YOUNG}

Section VI explores transitions of the {\it hexatic}
phase induced by the critical point. A modulated hexatic
phase is possible very near the critical point. If
there are no modulated hexatic instabilities, the coupling
of the Ising order parameter to the hexatic bond angle
field leaves the nontrivial Onsager-like critical behavior
of the 2D critical point intact.

\section{Free Energy Models}

In this section we describe the free energy functional for
the physics near an isostructural critical point in terms of
the elastic strain fields, an Ising-like structural order
parameter, and an interaction between these degrees of
freedom.  Dislocations, leading to multiple-valued
displacement fields will only  be considered in detail in
later sections.  Symmetries dictate the form of possible
coupling terms.

A first order transition in a scalar
quantity such as density near a
critical point can be described by an
Ising model. In the continuum
representation, the
``Landau-Ginzburg-Wilson'' expression
for the free energy takes the form
\cite{GOLDENFELD}

\begin{equation}
{\cal H}_{I} \equiv {F_{I} \over k_{B}T} =
 \int d^{2}x\, \left({1 \over 2} \vert\nabla \varphi
\vert^{2} + {r \over 2} \varphi ^{2}
+  {u \over 4!} \varphi^{4}
- h\varphi +\ldots\right),
\label{HISING}
\end{equation}

\noindent where for a first order
liquid-gas transition or isostructural
solid-solid transition, $\varphi$ is
defined as the density deviation from
the critical density; $\varphi =
\rho-\rho_{c}$. Near criticality, $u$ is
approximately independent of $T$ and $r
\simeq a _{2}(T-T_{c})/T_{c}$. The linear term
proportional to $h$ represents a
chemical potential and allows us to
study the transitions at densities away
from the critical isochore. Positive $h$
favors $\varphi > 0$ ({\it i.e.} the
dense phase $S_{2}$ in Fig. 1) while
$h<0$ favors $\varphi < 0$ (the dilute
solid phase $S_{1}$ in Fig.  1).  A
potential third order coupling,
($w\varphi^{3}$, say) in Equation
(\ref{HISING}) can be eliminated by a
shift in $\varphi$.

In addition to an Ising order parameter
describing the phase separation, a key feature
of the solid phases is their resistance to
deformations.  The elastic
deformation energy is \cite{LANDAU6},

\begin{equation}
{\cal H}_{el} \equiv {F_{el}\over k_{B}T} =
 {1 \over 2}\int d^{2}x\, \left(2\mu e_{ij}^{2}
+ \lambda e_{ii}^{2}\right) \label{Helastic}
\end{equation}

\noindent where $e_{ij} = {1 \over
2}(\partial_{i}u_{j} +
\partial_{j}u_{i})$ is the symmetric
strain tensor and $\mu$ and $\lambda$
are the Lam\'e coefficients defined as
the microscopic Lam\'e coefficients
divided by $k_{B}T$. Equation
(\ref{Helastic}) is valid for
two-dimensional triangular crystals
undergoing small strains.

The free energy (\ref{Helastic}) in general contains both
smoothly varying single-valued strains as well as
singular contributions from point-like defects such as
dislocations.  When a solid is thermally excited, defect
complexions must be explicitly included in the statistical
weight $e^{-{\cal H}_{el}}$, as well as in a coupling term,
such as Eq. (\ref{Hint}) below. As discussed in Sections IV
and V, dislocations can be incorporated by considering
the strains as comprised of a smooth ``phonon''
component, $u_{ij}$, and a singular part, $w_{ij}$,  due to
the presence of defects,  so that $e_{ij} = u_{ij} + w_{ij}$.

We now consider the coupling between
the order parameter $\varphi$ and the
strain.  Since the model is isotropic,
symmetry requires that to leading order, only the
dilatation, $e_{\ell\ell}$, couples to
the phase variable.
The lowest order interaction term for a
phase separating system is

\begin{equation}
{\cal H}_{int} = g \!\int \!d^{2}x\, \varphi\, e_{\ell \ell}.
\label{Hint}
\end{equation}

\noindent We assume $g>0$ so that a
lattice compression ($e_{\ell \ell} <
0$) will bias the system toward
$\varphi>0$, {\it i.e.} toward the {\it
denser} isostructural solid $S_{2}$ in
Figure \ref{SSPHASE}.  Related
couplings between  an elastic strain
field and an auxiliary field have been
studied in the context of compressible
statistical spin models
\cite{BERGMAN,SAK}, quenched random
impurities \cite{DRN83}, and
vacancy/interstitial defects in
crystals.

For compressible magnets with an Ising-like order
parameter,  the relevant interaction energy is
\cite{BERGMAN,SAK}

\begin{equation}
{\cal H}_{int} = g^{\prime}\!\int \! d^{2}x\,
\varphi^{2}\, e_{\ell \ell}.
\label{Hint2}
\end{equation}

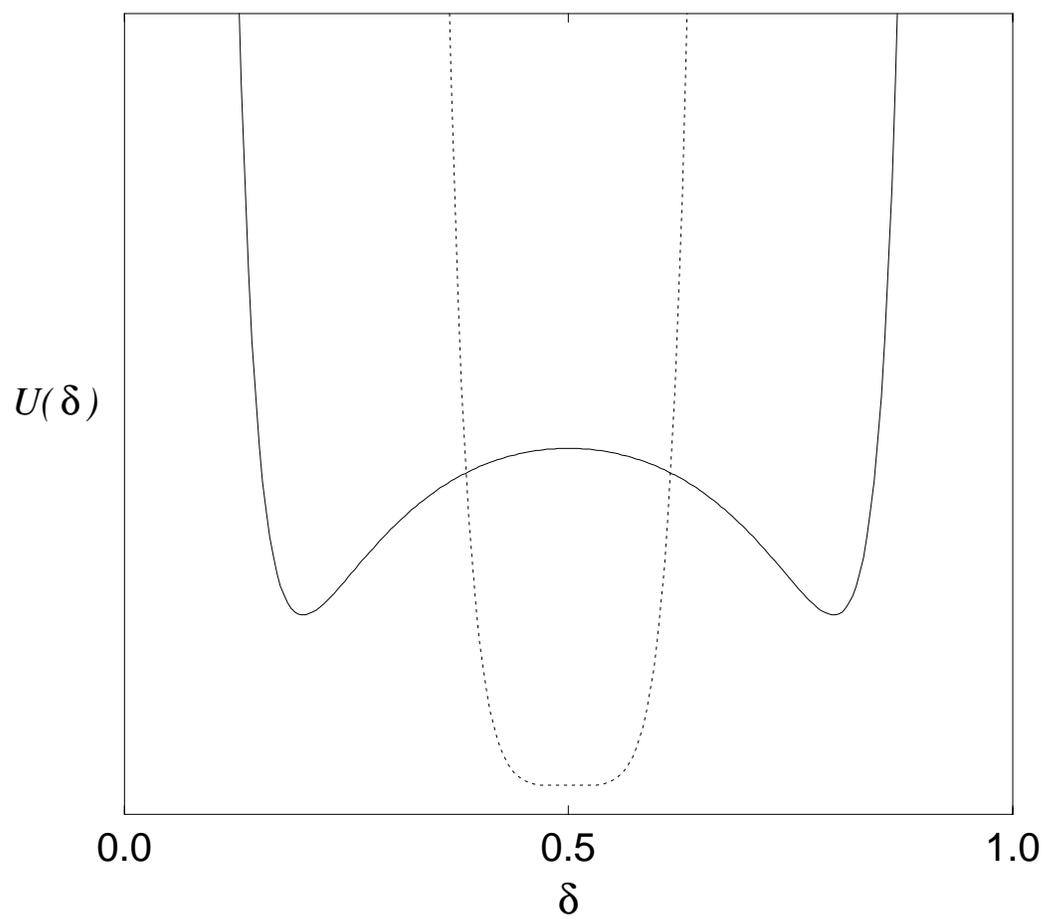
\begin{figure}
\caption{Interaction potential $U(\delta)$ between
particles and flat confining walls. The distance $\delta$
along the ordinate measures the distance away from the
mid-plane $\delta = 0.5$
between the walls. When the plates are closely
spaced $U(\delta)$ is given by the dashed curve
corresponding to an Ising free energy above critical
temperature. Increasing the plate gap lowers the
temperature of the Ising-like free energy (solid
curve).}\label{WP}
\end{figure}

\noindent A model of this kind would be appropriate for
colloidal crystals symmetrically confined between glass
plates with attractive wall potentials of the kind shown in
Figure \ref{WP}. Such a double wall potential can induce
(for the appropriate pair potentials between the colloidal
particles)  an Ising-like isostructural critical point inside
the two-dimensional crystalline phase. Below the Ising
phase transition, the crystal prefers to sit closer to one
wall or the other. If $\varphi(\vec{x})$ represents the
deviation of the local particle position from the
mid-plane, we can use Eq. (\ref{HISING}) (with $h=0$) to
represent the Ising degrees of freedom and Eq.
(\ref{Helastic}) to model the elastic ones. The symmetrical
confinement (we neglect gravity) insures that a coupling
of the form (\ref{Hint}) is impossible, so we are left with
(\ref{Hint2}).

As discussed, for example, by Bergman and Halperin
\cite{BERGMAN}, this coupling changes the critical
exponents of the Ising-like transition in the {\it absence}
of dislocations provided the specific heat exponent
$\alpha$ of the pure Ising system is positive.  Because
$\alpha=0$ for the 2D Ising model, there is at most a very
slow marginal crossover away from pure Ising-like
behavior for these symmetrically confined crystals. The
coupling $g^{\prime}$ induces a specific heat-like
singularity in the inverse elastic bulk modulus
$\bar{B}^{-1}$ near the transition.  The
renormalized inverse bulk modulus can be defined as

\begin{equation}
\begin{array}{ll}
\displaystyle  {1 \over \bar{B}} & \displaystyle ={1 \over A}
\langle U_{ii}U_{jj}\rangle \\
\displaystyle \: & \displaystyle = {1 \over B} +
{g^{\prime 2}\over B^{2}}{1 \over
A}\int d^{2}x\,d^{2}x^{\prime}\langle
\varphi^{2}(\vec{x})\varphi^{2}(\vec{x}^{\prime})\rangle
\simeq {1 \over B} + a_{2}{g^{\prime 2} \over
B^{2}}\left({T-T_{c}\over T_{c}}\right)^{-\alpha},
\label{Binv1}
\end{array}
\end{equation}

\noindent where $U_{ii} = \int d^{2}x u_{ii}(\vec{x})$, $A$
is the system area, and $B\equiv \mu+\lambda.$  If the
specific heat diverges, $\bar{B} = 0,$ the energy of one isolated
dislocation will become small, and we again expect
dislocation mediated melting close to the
critical point, similar to the argument in Ref.
\cite{FRENKEL1}.
However, this specific heat singularity is much weaker
than the singularity which induces melting for an
asymmetrical model, so a hexatic phase may be difficult
to observe experimentally.

In the remainder of  this paper, we consider only the
linear coupling, (\ref{Hint}) where the effects are expected
to be more dramatic. The total free energy is thus,

\begin{equation}
{\cal H} = \int d^{2}x\left[
 \mu e_{ij}^{2}+{\lambda
\over 2}e_{ii}^{2}\right]
 + g\int d^{2}x e_{\ell
\ell}\varphi + \int d^{2}x \left[
 \frac{1}{2}\vert\nabla
\varphi\vert^{2}+\frac{r}{2}\varphi^{2} +
\frac{u}{4!}\varphi^{4} -h \varphi +  \ldots \right].
\label{H}
\end{equation}

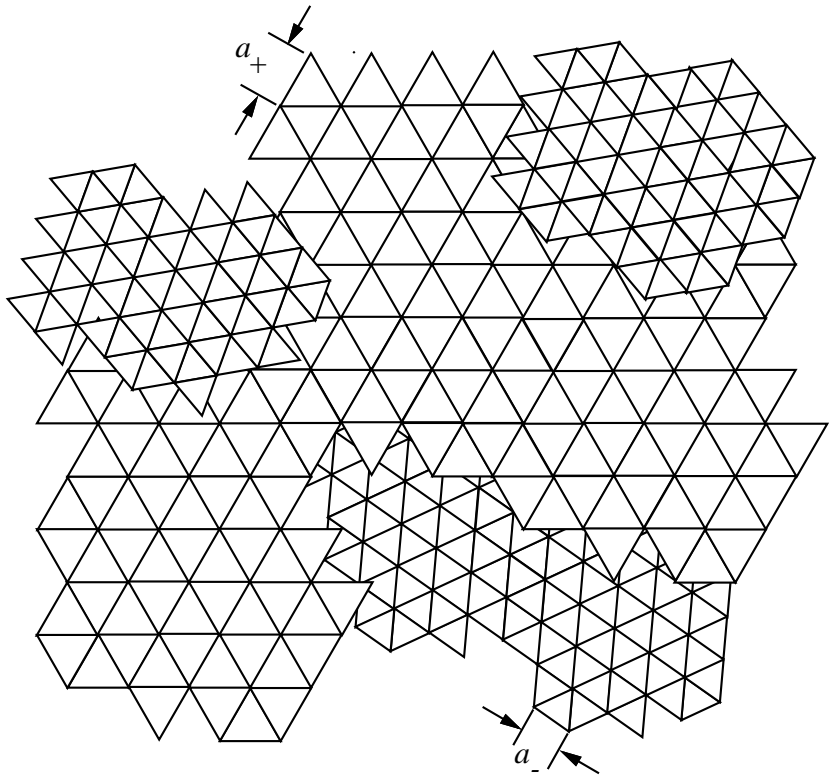
\begin{figure}
\caption{Coexistence of two isostructural solid phases,
$S_{1}$ and $S_{2}$ with lattice constants
$a_{+}$ and $a_{-}$respectively.}
\label{DISTORT}
\end{figure}

\noindent Figure \ref{DISTORT} shows schematically the
coexistence of two different density phases of lattice
constant $a_{+}$ and $a_{-}$.  Under a uniform area
change, the system responds by undergoing both pure
dilatational elastic deformations in each of the phases as
well as interconversion between local regions
of phase, $S_{1}$ or
$S_{2}$.  The relative partition between these two
responses is governed by $g$.  In Appendix A, we discuss
the stability constraints on the coupling constants
determined from Eq. (\ref{H}) expanded to quadratic order.

\section{Defect-Free Limit}

In this section, we study the interplay between the strain
and Ising fields in the absence of defects.  The strain
fields $u_{ij}$ are single-valued and can be thermally
averaged out in the quadratic theory. We assume free
boundary conditions in the $(x_{1}, x_{2})$
 - plane and  write the Fourier transformed
strain in terms of uniform bulk distortions and finite
wavevector modes \cite{LARKIN}

\begin{equation}
u_{ij}(\vec{x}) = u_{ij}^{(0)} + {1\over
2A}\sum_{\vec{q}\neq 0}
i(q_{i}u_{j}(\vec{q})+
q_{j}u_{i}(\vec{q}))e^{i\vec{q}
\cdot\vec{x}}. \label{DECOMP6}
\end{equation}

\noindent In the decomposition (\ref{DECOMP6}), the
uniform symmetric strains $u_{ij}^{(0)}$ must be treated
separately because these require {\it three} degrees of
freedom, in contrast to the two independent phonon
modes which arise when  $\vec{q}\neq 0$.  A similar
treatment is required in the study of the zero-wavevector
excitation in Bose-Einstein condensation.

Similar to (\ref{DECOMP6}), we decompose
$\varphi(\vec{x})$ as

\begin{equation}
\begin{array}{rl}
\displaystyle \varphi(\vec{x}) &
= \varphi_{0} + \sum_{\vec{q}\neq
0}\tilde{\varphi}(\vec{q})e^{i\vec{q}\cdot\vec{x}} \\[13pt]
\displaystyle \: & = \varphi_{0} + \delta\varphi(\vec{x}).
\end{array}
\end{equation}

\noindent For the linear coupling model
(\ref{Hint}) considered here, the zero
wavevector strain modes couple only to
the zero wavevector Ising field
$\varphi_{0}$.  After integrating out these
uniform bulk strains, we find an
effective energy for $\varphi_{0}.$
Averaging over the fluctuating crystal
displacements, $(\vec{u}(\vec{q}), \vec{q} \neq 0)$,
however, gives a different effective
energy for the nonzero wavevector Ising
fields $\delta\varphi(\vec{x})$. The
final result including the quartic coupling is

\begin{equation}
\begin{array}{r}
\displaystyle {\cal H}_{\em
eff}[\varphi]= A\left[
\left({r \over 2} -{g^{2}\over 2B}
\right)\varphi_{0}^{2}+
{u \over 4!}\varphi_{0}^{4} -
h\varphi_{0}\right] +
\displaystyle {1 \over 2}\!
\int \!d^{2}x \left[ \vert\vec{\nabla}
\varphi\vert^{2} +
\left(r- {g^{2}\over (\mu+B)}+
{u\over 2}\varphi_{0}^{2}\right)
\delta\varphi^{2}(\vec{x})\right] \\[13pt]
\displaystyle  + {u \over 6 A^{3}}
\varphi_{0}\sum_{\vec{q}_{1}\neq 0}
\sum_{\vec{q}_{2}\neq
-0,\vec{q}_{1}}\!
\varphi(\vec{q}_{1})
\varphi(\vec{q}_{2})\varphi
(-\vec{q}_{1}-\vec{q}_{2}) +
{u \over 4!}\!\int \! d^{2}x
\delta\varphi^{4}(\vec{x}) + \ldots \label{Heff2}
\end{array}
\end{equation}

\noindent where $\delta\varphi(\vec{x})$
contains only finite wavevector
perturbations of the Ising order
parameter about its uniform value
$\varphi_{0}$.

The $\vec{q}= 0$ strains
$(u_{ij}^{(0)})$  renormalize the $\varphi_{0}$ modes
differently than the finite wavevector
phonons $(\vec{u}(\vec{q}))$,
renormalize the $\varphi(\vec{q}\neq 0)$
modes. The critical temperature of
$\varphi_{0}$ is shifted according to $T^{*}
\approx T_{c}(1+g^{2}/a_{2}B)$.  The
mass or quadratic coefficient of the
$\varphi_{0}$ field (denoted $\bar{r}_{0}$), will, upon
decreasing temperature, always vanish
before the quadratic coupling of the
finite wavevector modes, (denoted $\bar{r}$).
Thus, $\bar{r}_{0} < \bar{r}$, where

\begin{equation}
\bar{r}_{0} \equiv
r -{g^{2}\over B}
\label{rbar0}
\end{equation}

\noindent and,

\begin{equation}
\bar{r} \equiv
r  - {g^{2} \over (\mu+B)} + {u
\over 2}\varphi_{0}^{2}.
\label{rbar}
\end{equation}

Since the  $\delta\varphi(\vec{x})$ modes remain massive
near $\bar{r}_{0} \approx 0$, they can be integrated out
in perturbation theory to give a renormalized effective
potential (with no gradient term) for the remaining
$\varphi_{0}$ degree of freedom. It is easily checked that
the ordering of $\varphi_{0}$ prevents the subsequent
ordering of the $\delta\varphi(\vec{x})$ modes at a lower
temperature within our model.  We conclude that the
critical behavior of a 2D Ising model
interacting with strain fields according to
Eq. (\ref{Hint}) is described by
{\it mean-field} exponents.

The physical reason for mean-field critical behavior is
that the uniform strains induce an infinite-ranged
interaction between the zero wavevector Ising degrees of
freedom. To see this, note that the quadratic
contributions to Eq. (\ref{Heff2}) may be combined to give

\begin{equation}
{\cal H}_{\em eff}^{(2)} = {1 \over 2}\int d^{2}x\left[ \vert
\nabla \varphi(\vec{x})\vert^{2} + \left(r - {g^{2} \over
\mu+B}\right)\varphi^{2}(\vec{x})\right] \\ \nonumber
- {\mu g^{2} \over A(\mu + B)B}\int \! d^{2}x\!\int
\!d^{2}x^{\prime} \varphi(\vec{x})
\varphi(\vec{x}^{\prime}).
\end{equation}

\noindent The first term is the usual continuum
representation of a short-ranged interaction, but the
second has infinite range, induced by the strain coupling.

It is also instructive to consider elastic constants as
renormalized by fluctuations of the Ising variable.
It is easy to show that integrating over the slowly
varying $\varphi$ degrees of freedom
leads to  an effective Lam\'{e} coefficient

\begin{equation}
\bar{\lambda} = \lambda - g^{2}/r^{\prime},
\label{EFFL1}
\end{equation}

\noindent where

\begin{equation}
r^{\prime} = r + {u \over 2}\varphi_{0}^{2}.
\label{rprime}
\end{equation}

Consider the behavior of the effective elastic constant
$\bar{\lambda}$ near the (mean field) critical
temperature $T^{*}$ discussed above.  Recall that
$\bar{r}_{0}(T^{*}) = 0$, and let us assume for
simplicity that $h=0$ and $T \leq T^{*}$. Since
$r^{\prime} = r$ in this case, we have
$r^{\prime}(T^{*}) = r(T^{*}) = g^{2}/B =
g^{2}/(\mu+\lambda)$. It follows that

\begin{equation}
\bar{\lambda}(T^{*}) = -\mu
\end{equation}

\noindent so that the effective bulk modulus

\begin{equation}
\bar{B}(T) = \mu + \bar{\lambda}
\end{equation}

\noindent vanishes as $T\rightarrow T^{*}.$ This
vanishing of the effective bulk modulus is expected
quite generally, since $\bar{B}$ is related to the
Ising fluctuations in our model by an equation similar
to (\ref{Binv1}), namely,

\begin{equation}
\begin{array}{ll}
\displaystyle {1 \over \bar{B}} & \displaystyle = {1 \over A}\langle
U_{ii}U_{jj}\rangle \\
\displaystyle \: & \displaystyle ={ 1\over B} + {g^{2} \over B^{2}}{1
\over A}\int d^{2}x \int d^{2}x^{\prime} \langle
\varphi(\vec{x})\varphi(\vec{x}^{\prime})\rangle.
\label{Binv2}
\end{array}
\end{equation}

\noindent The strongly divergent mean field Ising fluctuations,
${1 \over A}\int d^{2}x\int d^{2}x^{\prime}\langle
\varphi(\vec{x})\varphi(\vec{x}^{\prime})\rangle \sim
(T-T^{*})^{-1}$, insure that $\bar{B}$ vanishes.
The reduction in $\lambda$ implied by
(\ref{EFFL1}) corresponds to a lattice
softening, suggesting that dislocations can more easily
form.

Linear stability analysis of the defect-free model is given
in Appendix A. The conditions of stability are different for
zero wavevector and finite wave vector instabilities
consistent with the renormalized ``masses''  for
$\vec{q}= 0$ and $\vec{q}\neq 0$ fluctuations,
$\bar{r}_{0}$ and $\bar{r}$, respectively.

\section{Isolated Dislocation}

In this Section, we determine the
behavior of $\varphi$ and
 the strains with a single
dislocation at the origin. We also
calculate the free energy of this
dislocation and apply an energy-entropy
balance to predict when dislocations are
favored. The dislocation contribution to
the field configurations is found by
solving the extremal equations developed
below.

We first write the total strain as

\begin{equation}
e_{ij} = u_{ij}^{(0)} + w_{ij}
\end{equation}

\noindent where $u_{ij}^{(0)}$ is the $\vec{q}
= 0$ part and $w_{ij}$ are the singular strains due to
the dislocation. After minimizing with respect to
 $u_{ij}^{(0)}$ and expanding
$\varphi(\vec{x})$ about a uniform value $\varphi_{0} +
\delta\varphi(\vec{x})$, we find that the effective free
energy becomes (to quadratic order in $\delta\varphi$)

\begin{equation}
{\cal H}_{{\em eff}}[\varphi,w_{ij}] =
A\left[ {\bar{r}_{0} \over 2}
\varphi_{0}^{2} + {u \over 4!}
\varphi_{0}^{4} - h\varphi_{0}\right]
 + \int d^{2}x\,\left[
\mu w_{ij}^{2} + {\lambda
\over 2}w_{ii}^{2} +
g\delta\varphi w_{ii} + {1 \over 2}
\vert\nabla\delta\varphi\vert^{2}
+ {r^{\prime} \over 2}
\delta\varphi^{2} + O(\delta\varphi^{3})
\right],
\label{Heff}
\end{equation}

\noindent where $\bar{r}_{0}$ is given by Eq.
(\ref{rbar0}) and $r^{\prime}$ is given by Eq.
(\ref{rprime}). Near a dislocation $\varphi$
distorts from $\varphi_{0}$.  We will assume
that $\varphi(\vec{x}\rightarrow \infty)
\rightarrow \varphi_{0}$,
where $\varphi_{0}$ is one of possibly two uniform
minima. This {\it ansatz} is appropriate  in the presence of
a single dislocation, since domain walls connecting
regions with $\varphi = \varphi_{0}$ and $\varphi =
-\varphi_{0}$ would cost an additional gradient energy
scaling as the system size.

After minimizing (\ref{Heff}) with respect to
$\varphi_{0}$, $\delta\varphi(\vec{x})$, and $\vec{u}$,
the displacement caused by the dislocation, we obtain

\begin{equation}
\bar{r}_{0} \varphi_{0} + {u \over 6}\varphi_{0}^{3}-h
 +{u \over 2}\varphi_{0}
\int d^{2}x \delta\varphi^{2}(\vec{x})= 0,
\label{EQM1}
\end{equation}

\begin{equation}
-\nabla^{2}\delta\varphi +
r^{\prime}\delta\varphi + gw_{ii} +
O\left(\delta\varphi^{2}\right) = 0,\label{EQM2}
\end{equation}

\noindent and,

\begin{equation}
\mu\nabla^{2}\vec{u}+(\mu+\lambda)\vec{\nabla}(\vec
\nabla\cdot\vec{u}) + g \vec{\nabla}\delta\varphi =
-\mu \hat{z}\times\vec{b}\delta(\vec{x}),
\label{EQM3}
\end{equation}

\noindent where $\vec{u}$ in the last equation is
understood to be the lattice displacement due to the
dislocation, $w_{ij} = {1 \over
2}(\partial_{i}u_{j}+\partial_{j}u_{i})$. These strains which
satisfy the extremal equations guarantee vanishing of
cross terms between $u_{ij}$ and $w_{ij}$ in (\ref{Heff}).
We the treatment in Ref. \cite{LANDAU6} and
write the displacement vector as
$\vec{u} = \vec{u}_{0} + \vec{u}^{\prime}$, where
$\vec{u}_{0}$ assumes the Burger's vector when integrated around
a dislocation, {\it i.e.}, $\nabla^{2}\vec{u}_{0} = -\hat{z}\times\vec{b}
\delta(\vec{x}).$ Note also that $\nabla\!\cdot\!\vec{u}_{0} = 0$.
By taking the divergence of (\ref{EQM3}) and substituting in
the expression for $(\vec{\nabla}\cdot\vec{u})$ from
(\ref{EQM2}), we obtain a closed equation for
$\delta\varphi(\vec{x})$,

\begin{equation}
-\nabla^{4}\delta\varphi +
\left(r^{\prime}- {g^{2} \over
\mu+B}\right)\nabla^{2}\delta\varphi
= {2\mu g \over \mu+B}
\nabla\cdot\hat{z}\times\vec{b}\delta(\vec{x}).
\end{equation}

Note that we have retained the gradient contribution to
the energy in Eq.  (\ref{EQM2}).  The value of
$\varphi_{0}$ must be determined self-consistently;
however, for small $\delta\varphi(\vec{x})$, the
correction to $\bar{r}_{0}$ in (\ref{EQM1}) is small, and
$\varphi_{0}$ can be approximated by the solution to

\begin{equation}
\bar{r}_{0}\varphi_{0} + {u \over 6}\varphi_{0}^{3} - h
\approx 0. \label{MIN}
\end{equation}

Having determined the boundary
conditions for a single dislocation at
the origin, we can solve for a small
deviation of the Ising order parameter
$\delta\varphi(\vec{x})$. With $\vec{x} =
\{x_{1}, x_{2}\}$, and $\vec{b} =
b\hat{e}_{1}$, where $\hat{e}$ is a unit
vector in the $x_{1}$ direction,

\begin{equation}
\delta\varphi(\vec{x})  \simeq
{\mu g b \over \pi \kappa^{2} (\mu+B)}
{x_{2} \over
\vert\vec{x}\vert^{2}}\left[1 -
\sqrt{{\pi \over 2\kappa
\vert\vec{x}\vert}}
e^{-\kappa
\vert\vec{x}\vert}\right]
\label{deltavarphi}.
\end{equation}

\noindent where

\begin{equation}
\begin{array}{rl}
\displaystyle\kappa^{2} & = \bar{r} \\
\: & \displaystyle = r^{\prime} - {g^{2}\over \mu+B}.
\label{KAPPA}
\end{array}
\end{equation}

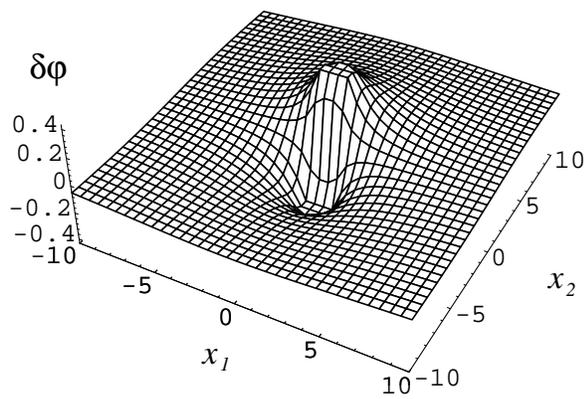
\begin{figure}
\caption{The solution $\delta\varphi$ for $\vec{b} =
-b\hat{e}_{1}$.  Note that $\varphi$ diverges at the
dislocation, and a cutoff must be imposed before
$\delta\varphi$ gets too large;  however, unlike the
phonon displacement field $\vec{u},
\delta\varphi(\vec{x})$ is single-valued.}\label{varphi}
\end{figure}

\noindent A plot of $\delta\varphi(\vec{x})$ is shown in
Figure \ref{varphi}. The peaks and valleys ($x_{2}>0$ and
$x_{1}<0$ respectively) correspond to a missing or extra
row of particles.  We  also calculate the dilatation,
Eq. (\ref{EQM2});

\begin{equation}
\nabla\!\cdot\!\vec{u} = w_{ii} \simeq -{\mu b r^{\prime}\over
\pi\kappa^{2}(\mu+B)}{x_{2} \over
\vert\vec{x}\vert^{2}}
\left[1- O\left(\vert\vec{x}\vert^{3}
e^{-\kappa \vert\vec{x}\vert}\right)
\right]\label{Wll}
\end{equation}

\noindent which is asymptotically inversely proportional
to $\delta\varphi$. The total displacements can be found from

\begin{equation}
\mu\nabla^{2}\vec{u}^{\prime} = -(\mu+\lambda)\vec{\nabla}w_{ii}-
g\vec{\nabla}\delta\varphi - 2\mu b\delta(\vec{x})\hat{e}_{2}.
\end{equation}

\noindent We note that $\partial w_{ii}/\partial x_{2}$
and $\partial \delta\varphi/\partial x_{2}$ behave as
$\delta$-functions at the origin, and find the total displacements
around a dislocation,

\begin{equation}
\begin{array}{l}
\displaystyle u_{1}= {b \over 2\pi}\left[ \tan^{-1}
\left({x_{2} \over x_{1}}\right)+{Br^{\prime}
-g^{2} \over \kappa^{2}(\mu+B)}{x_{1}x_{2}
\over \vert\vec{x}\vert^{2}}\right]
+O(e^{-\kappa\vert\vec{x}\vert}) \\[13pt]
\displaystyle u_{2} = -{b \over 2\pi}
\left[{\mu r^{\prime} \over \kappa^{2}(\mu+B)}\ell n \vert\vec{x}\vert
+ {Br^{\prime} -g^{2} \over \kappa^{2}(\mu+B)}
 {x_{1}^{2}\over \vert\vec{x}\vert^{2}}\right]
+O(e^{-\kappa\vert\vec{x}\vert})
\label{DISPLACEMENTS}
\end{array}
\end{equation}

\noindent These displacements are
asymptotically identical to those of
standard elasticity theory \cite{LANDAU6}, except with
$\lambda$ replaced by $\bar{\lambda}$.

Evaluating (\ref{Heff}) to quadratic order using
asymptotic expressions for  $\vec{u}$ and $\delta\varphi$
gives the total energy of a single dislocation of
strength $b$:

\begin{equation}
{\cal H}_{{\em eff}} =
{\bar{K}b^{2}\over 8\pi}\ell n\left({R
\over a}\right) + c(a),
\label{DE}
\end{equation}

\noindent where $R$ is the linear sample dimension
and $c(a)$ is function of the dislocation core size
$a$. The most singular part of the energy
(\ref{DE}) has the same form as the standard result
except with effects of the coupling ${\cal H}_{int}$
contained in an effective Young's modulus

\begin{equation}
\bar{K} = {4\mu(\mu+\bar{\lambda}) \over
2\mu+\bar{\lambda}}\label{Kbar}
\end{equation}

\noindent where the effective Lam\'{e}
coefficient for a dislocation is given by

\begin{equation}
\bar{\lambda} = \lambda - {g^{2} \over r^{\prime}},
\label{EFFL2}
\end{equation}

\noindent consistent with Eq. (\ref{EFFL1}). As
discussed in Section III, this reduction of $\lambda$
causes the effective bulk modulus $\bar{B} =
\mu+\bar{\lambda}$ to vanish near the isostructural
critical point.

The total free energy of a dislocation, $G_{1}$,
also includes entropy describing configurations
associated with the  $(R/a)^{2}$ possible
positions the dislocation can occupy. Hence,

\begin{equation}
G_{1}/k_{B}T = {\cal H}_{{\em eff}}
 - 2\ell n(R/a).
\end{equation}

\noindent The temperature above which the entropy
favors dislocation formation is given by $G_{1}/k_{B}T
\approx 0$, or,

\begin{equation}
\bar{K} \approx 16\pi. \label{KT1}
\end{equation}

\noindent This criterion, modified from the
Kosterlitz-Thouless criterion of the decoupled theory
($4\mu B/(\mu+B)
\approx 16\pi$) determines the phase diagram.  The
onset of dislocation melting occurs when $K > \bar{K}$,
where

\begin{equation}
\begin{array}{rl}
\displaystyle \bar{K} & \displaystyle =
{4\mu\left[B r^{\prime}-g^{2}\right]
\over (\mu+B)r^{\prime}-g^{2}} \\[13pt]
\displaystyle \: & \displaystyle = {4\mu(\mu+\bar{\lambda}) \over
2\mu + \bar{\lambda}} = 16\pi.\label{Kbar2}
\end{array}
\end{equation}

\noindent This condition is equivalent to the usual
Kosterlitz-Thouless criterion, with the replacement
$B \rightarrow \bar{B} = B - g^{2}/r^{\prime}.$ Because
$\bar{B}(T)\rightarrow 0$ as $T \rightarrow T^{*}$,
melting is inevitable near the isostructural critical
point. To
obtain phase diagrams, we assume $\mu$ and $B$ to be
slowly varying functions of the order parameter
$\varphi$. If we further assume for simplicity that bare
elastic constants (unrenormalized by either Ising order
parameter fluctuations or dislocations) satisfy
$\mu(\rho_{c}) \approx \lambda(\rho_{c})$, the original
dislocation mediated melting theory ($g=0$), predicts dislocation
unbinding when $\mu(\rho_{c}) , \lambda(\rho_{c})
\approx 6\pi.$ However, in the modified
Kosterlitz-Thouless criterion, (\ref{Kbar2}), the effective
parameters may be such that (\ref{Kbar2}) is satisfied even
when $\mu,\lambda > 6\pi$.  In the ($\bar{r}_{0}/u,
\rho-\rho_{c}$) plane, the boundary of a hexatic phase
near the isostructural critical point as
 found from (\ref{Kbar2}) is shown in Figure
\ref{SSPHASE}.

We use  the solution of the cubic equation (\ref{MIN}) in
(\ref{Kbar2}) to obtain a quantitative picture in the $h-T$
plane.   Phase boundaries for $\mu= \lambda = 10\pi$ in
the $\bar{r}_{0}/u$ and $h/u$ plane are plotted in Figure \ref{hT1}.
More precise determination of the phase boundary would
require a first principles calculation of the
unrenormalized elastic moduli $\mu$ and $\lambda$.
This can be obtained for example by density functional
theory \cite{TAREYEVA}  or primitive cell models
\cite{FRENKEL0}. Our simple model however, shows
salient features of  2D melting induced by an
isostructural critical point and agrees qualitatively with
the simulated phase diagram of Bladon and Frenkel
\cite{FRENKEL1}.

\section{Effects of Many Dislocations}

When many dislocations are present, the possibility of
more complicated structures arises. In particular,
dislocations may aggregate to form grain
boundaries, or extended domains
where $\varphi$ may take on values different from
$\varphi_{0}$, or for $\bar{r}_{0}< 0$, both domains of
$+\varphi_{0}$ and $-\varphi_{0}$ phases may coexist.
We will assume these complex aggregates do not form. Here, we
assume a dilute dislocation gas, and demonstrate that
the renormalization group recursion relations for
the effective parameters are unchanged from those
of the uncoupled melting theory.

The stress tensor associated with
${\cal H} \equiv {\cal H}_{e\ell}
+{\cal H}_{I} + {\cal H}_{int}$ is

\begin{equation}
\sigma_{ij} =
(\mu+\lambda)e_{\ell \ell}\delta_{ij}
+ 2\mu(e_{ij}-{1\over 2}
e_{\ell \ell}\delta_{ij})+2g\varphi\,\delta_{ij},
\end{equation}

\noindent which can be inverted to give

\begin{equation}
e_{ij} = {1 \over 2\mu}\sigma_{ij} -{\lambda \over
4\mu(\mu+\lambda)}\sigma_{\ell \ell}\delta_{ij} - {g
\over
\mu+\lambda}\varphi\,\delta_{ij}.\label{STRAIN}
\end{equation}

\noindent Functional minimization with respect to
the phonon displacements
yields a divergence-free stress,

\begin{equation}
\partial_{i}\sigma_{ij}^{(s)} =
0\label{DIV}
\end{equation}

\noindent to which must be added single-valued
fluctuating stresses $\tilde{\sigma}_{ij}$. Since (\ref{DIV})
holds away from defects, the extremal stress can be
expressed in terms of a double curl, \cite{LANDAU6}

\begin{equation}
\sigma_{ij}^{(s)} \equiv
\epsilon_{im}\epsilon_{jn}
\partial_{m}\partial_{n}\chi.
\end{equation}

\noindent For the present problem, the stress
tensor for a collection of dislocations at
 positions $\{\vec{x}_{\alpha}\}$
with Burgers' vectors $\{\vec{b}_{\alpha}\}$ and
``Burger's charge density'' $\vec{b}(\vec{x}) =
\sum_{\alpha}\vec{b}_{\alpha}\delta(\vec{x}-\vec{x}_{\alpha})$
is given by

\begin{equation}
\begin{array}{ll}
\displaystyle \sigma_{ij} & = \displaystyle
\sigma_{ij}^{(s)} + \tilde{\sigma}_{ij} \\
\displaystyle \: & \displaystyle =-{K_{0} \over
4\pi}\epsilon_{im}\epsilon_{jn}\partial_{m}\partial_{n}
\int d^{2}x^{\,\prime}
b_{k}(x^{\,\prime})\epsilon_{kl}(x-x^{\,\prime})_{\ell}
\left[\ell n {\vert \vec{x}-\vec{x}^{\,\prime}\vert \over a}
+ C\right] \nonumber \\
\: & \displaystyle + {gK_{0} \over \mu+\lambda} \epsilon_{im}
\epsilon_{jn}\partial_{n}\partial_{m}\int
d^{2}x^{\,\prime}
G(x-x^{\,\prime})\nabla^2\varphi(x^{\,\prime}) +
\tilde{\sigma}_{ij}
\end{array}
\end{equation}

\noindent where $\nabla^{4}G(\vec{x}) =
\delta(\vec{x})$ and $C$
is a constant of order unity. The functional form
of the free energy ${\cal H} = {1 \over 2}\int
d^{2}x \sigma_{ij}e_{ij} + {\cal H}_{I}$ becomes

\begin{equation}
{\cal H} = {\cal H}_{I}[\varphi]+
{\cal H}_{0}[u_{ij}] +
{\cal H}_{int}[\varphi, u_{ij}] + {\cal H}_{D}[\vec{b}]
{\cal H}_{int}^{sing}[\varphi, \vec{b}]\label{Htot}
\end{equation}

\noindent where $u_{ij}$ is the smoothly varying part of the strain and

\begin{equation}
{\cal H}_{D} = -{K_{0} \over 8\pi}
\sum_{\alpha\neq\beta}
\left[\vec{b}(\vec{x}_{\alpha})\cdot
\vec{b}(\vec{x}_{\beta})
ln\left({\vert \vec{x}_{\alpha}-\vec{x}_{\beta}
\vert \over a}\right) -
{\vec{b}(\vec{x}_{\alpha})\cdot(\vec{x}_{\alpha}-\vec{x}_{\beta})
\vec{b}(\vec{x}_{\beta})\cdot(\vec{x}_{\alpha}-\vec{x}_{\beta})
\over \vert\vec{x}_{\alpha}-\vec{x}_{\beta}\vert^{2}}\right] +
{E_{c} \over k_{B}T}\sum_{\alpha}\vert
\vec{b}(\vec{x}_{\alpha})\vert^{2}. \label{HD}
\end{equation}

\noindent The dislocation contribution to the
strain, $w_{ii}$, may be written in terms of
Burgers' vectors \cite{DRN78,DRN79,DRN83},

\begin{equation}
w_{ii}(\vec{x}) = -{K_{0} \over 4\pi (\mu+\lambda)}
\sum_{\alpha} { \hat{z}\!\cdot\!\vec{b}(\vec{x}_{\alpha})
\times (\vec{x}-\vec{x}_{\alpha}) \over \vert
\vec{x}-\vec{x}_{\alpha}\vert^{2}},
\end{equation}

\noindent which leads to

\begin{equation}
\begin{array}{ll}
\displaystyle {\cal H}_{int}^{sing}[\varphi, \vec{b}]  &
\displaystyle =
g\!\int \!d^{2}x w_{ii}\varphi \\ \nonumber
\displaystyle \: & \displaystyle =
{g K_{0} \over 4\pi(\mu
+\lambda)}\sum_{\alpha}\int d^{2}x\,
\varphi(\vec{x})
{\hat{z}\!\cdot\!\vec{b}(\vec{x}_{\alpha})
\!\times\!(\vec{x}-\vec{x}_{\alpha})
\over \vert \vec{x}-\vec{x}_{\alpha}\vert^{2}}.
\end{array}
\end{equation}

\noindent If the fluctuations in $\delta\varphi(\vec{x})$ are now
averaged out, the effective model of interacting
dislocations takes the form of ${\cal H}_{D}$, except with
the replacement $K_{0} \rightarrow \bar{K}$, with $\bar{K}$ given by
(\ref{Kbar2}). Although we have omitted the details,
these results follow from
straightforward generalizations of
(\ref{EQM1}),
(\ref{EQM2}), and (\ref{EQM3}). For example, (\ref{EQM3}) is
generalized as

\begin{equation}
\mu\nabla^{2}\vec{u}
+(\mu+\lambda)\vec{\nabla}(\vec{\nabla}\cdot\vec{u})
+g\vec{\nabla}\varphi =
\mu\sum_{\alpha}\hat{z}\times\vec{b}_{\alpha}\delta
(\vec{x}-\vec{x}_{\alpha}).
\end{equation}

The energy (\ref{Htot}) can be analyzed by rescaling dislocation core
sizes and distances in a way similar to that discussed in
\cite{YOUNG}. Although recursion relations for $g$ and
$r$ as well as the elastic coefficients can be found this
way, it is simpler to construct recursion relations
in terms of the effective coupling $\bar{K}$.
These recursion relations are
\cite{DRN79,YOUNG}

\begin{equation}
{d y(\ell) \over d \ell} = \left[ 2 - {\bar{K}(\ell) \over
8\pi}\right] y(\ell) + 2\pi y^{2}(\ell)e^{\bar{K}(\ell)/16\pi}
I_{0}(\bar{K}(\ell)/8\pi) + O\left(y^{3}(\ell)\right)
\label{RECURY}
\end{equation}

\begin{equation}
{d \bar{K}^{-1}(\ell) \over d \ell} = {3\pi \over 2} y^{2}(\ell)
e^{\bar{K}(\ell) /8\pi}\left[ I_{0}(\bar{K}(\ell)/8\pi)-
{1 \over 2} I_{1}(\bar{K}(\ell)/8 \pi)\right]
+O\left(y^{3}(\ell)\right)
\end{equation}

\noindent where $\bar{K}(\ell)$ is defined by (\ref{Kbar2}) and
$y(\ell)$ is the dislocation fugacity $y \simeq
e^{-E_{c}/k_{B}T}$. $\bar{K}(\ell)$ and $y(\ell)$ are the usual
scale-dependent  renormalization group coupling constants.
The $O(y^{2})$ term on the
right-hand-side of (\ref{RECURY}) is found by
considering three dislocation configurations
\cite{DRN79}.  Flows of the parameters, as one considers
increasing length scales $\ell$, which have $y\rightarrow
0$ indicate states with proliferated dislocations, or a
hexatic phase.  A melting criterion identical  with the
modified Kosterlitz-Thouless criterion, Equation
(\ref{Kbar2}), arises from analysis of (\ref{RECURY}).
Moreover, we conclude that the critical properties of
standard dislocation melting theory \cite{DRN79}, such
as an essential singularity in the specific heat, and a
divergence of the translational correlation length of the
form  $\xi_{+}(T) \sim exp\,(const./
\vert T-T_{m}\vert^{0.36963})$, are unchanged
from the result of the usual dislocation-mediated theory.

\section{Hexatic Melting and Disclination Unbinding}

In this Section, we examine the
stability of the critical-point-induced
hexatic phase.  Within the hexatic
phase, when many dislocations are
unbound, the free energy is described by
slow spatial variations in the bond angle
field \cite{DRN78,DRN79}. The lowest order
symmetry allowed coupling between the
bond angle $\theta$ and Ising order
parameter has the form  $v
\nabla\theta\!\cdot\!\nabla\varphi$, and
the total free energy is

\begin{equation}
{\cal H}_{hex} = {K_{A} \over 2} \int d^{2}x\,\vert
\nabla \theta\vert^{2} + v\int d^{2}x\,
\nabla\theta\!\cdot\!\nabla\varphi + {\cal H}_{I}[\varphi],
\label{HHEX}
\end{equation}

\noindent where $K_{A}$ is the hexatic stiffness constant;
$K_{A}(T)\rightarrow \infty$ as the freezing transition is approached
from the hexatic phase \cite{DRN79}. We now consider the behavior
of the bond angle and density order parameter in the presence of
disclinations. The hexatic phase predicted from the
standard dislocation melting theory will melt into an isotropic
fluid phase upon further increase in temperature.
Proliferating disclinations cause this first order transition.

For our model, (\ref{HHEX}),  we find that the
interactions between disclinations and
$\varphi$ fields decouple. To see this, we first
write the extremal equations
derived from (\ref{HHEX});

\begin{equation}
K_{A}\nabla^{2}\theta + v\nabla^{2}\varphi = 0
\label{EQMHEX1}
\end{equation}

\begin{equation}
-\nabla^{2}\varphi + r\varphi + {u \over 6}\varphi^{3}-
h-v\nabla^{2}\theta = 0.\label{EQMHEX2}
\end{equation}

\noindent Note that Eq.
(\ref{EQMHEX1}) allows us to define a conjugate bond angle field

\begin{equation}
\epsilon_{ij}\partial_{j}\tilde{\theta}
\equiv \partial_{i}\theta + {v \over
K_{A}}\partial_{i}\varphi .
\end{equation}

\noindent The density order parameter is assumed to be
single-valued; however, the bond angle $\theta$ obeys
the noncommutivity relation implied by a set of disclinations
located at positions $\{\vec{x}_{\alpha}\}$,

\begin{equation}
\epsilon_{ij}\partial_{i}\partial_{j}\theta
= m(\vec{x}) = {\pi \over
3}\sum_{\alpha} s_{\alpha}\delta(\vec{x} -
\vec{x}_{\alpha}).
\end{equation}

\noindent The
strength of the $\alpha^{th}$
singularity is measured by the charge
$s_{\alpha}=\pm 1$.  Thus, we find

\begin{equation}
\partial_{i}\theta(\vec{x}) = \epsilon_{ij}
\int d^{2}x^{\prime}\,m(\vec{x}^{\prime})\partial_{j}
G(\vec{x}-\vec{x}^{\prime})- {v \over
K_{A}}\partial_{i}\varphi(\vec{x})
\label{Dtheta}
\end{equation}

\noindent Using Eq. (\ref{Dtheta}), we find that the terms coupling
the disclination positions to $\varphi(\vec{x})$ vanish and

\begin{eqnarray}
{\cal H}_{V} = {-\pi K_{A} \over
36}\sum_{\vec{x}\neq\vec{x}^{\prime}}
s(\vec{x})s(\vec{x}^{\prime})\ell n {\vert
\vec{x}-\vec{x}^{\prime} \vert \over a} + {E_{c} \over
k_{B}T}\sum_{\vec{x}}s^{2}(\vec{x}) + {K_{A} \over 2}\int
d^{2}x\, \vert \nabla \tilde{\theta}\vert^{2} \nonumber\\
+ v\int d^{2}x \nabla\tilde{\theta}\cdot\nabla\varphi +
 \int d^{2}x\,\left[ {1 \over 2}\vert\nabla\varphi\vert^{2} +
{r\over 2}\varphi^{2}
+ {u\over 4!}\varphi^{4} -
h\varphi\right],
\label{HV1}
\end{eqnarray}

\noindent where $\tilde{\theta}$ represents the smoothly
varying part of the bond angle field. In addition to a
possible disclination unbinding transition, a finite
wavevector instability may develop when higher order
terms in $\vec{\nabla}\varphi$ are included in
(\ref{HV1}). If we expand  (\ref{HV1}) to
quadratic order about the minimum of $\varphi$ and
average over the smooth density fluctuations, we find

\begin{eqnarray}
{\cal H}_{V} = {-\pi K_{A} \over
36}\sum_{\vec{x}\neq\vec{x}^{\prime}}
s(\vec{x})s(\vec{x}^{\,\prime})\ell n {\vert
\vec{x}-\vec{x}^{\prime} \vert \over a} + {E_{c} \over
k_{B}T}\sum_{\vec{x}}s^{2}(\vec{x}) + {K_{A} \over 2}\int
d^{2}x\, \vert \nabla \tilde{\theta}\vert^{2} \nonumber \\
+ {\bar{K}_{B}
\over 2} \int d^{2}x\, \vert\nabla^{2}\tilde{\theta}\vert^{2} +
{K_{C} \over 2}\int d^{2}x\, \vert \nabla
\nabla^{2}\tilde{\theta}\vert^{2} \label{HV2}
\end{eqnarray}

\noindent where we have added higher
order terms
${1 \over 2}K_{B}\vert\nabla^{2}\tilde{\theta}\vert^{2}
$ and ${1 \over 2}K_{C}\vert
\nabla\nabla^{2}\tilde{\theta}\vert^{2}$ to Eq. (\ref{HV1}),
and defined

\begin{equation}
\bar{K}_{B} \equiv  K_{B} - {v^{2}\over
r+{u\over 2}\varphi_{0}^{2}}.\label{KBbar}
\end{equation}

\noindent Either expression, (\ref{HV1})
or (\ref{HV2}), shows that the
interaction energy in (\ref{HHEX})
decouples from the disclination
configurations and only modifies the
effective stiffness of the $\varphi$
or $\theta_{0}$ fields.

Note from Eq. (\ref{KBbar}) that $\bar{K}_{B}$ becomes
negative near the isostructural critical point at $r=\varphi_{0} =
0$. For $\bar{K}_{B} < 0$, a finite
wavevector instability in bond angle
(along with density) occurs at

\begin{equation}
q^{*} = \pm {\vert \bar{K}_{B}\vert
 \over
\sqrt{3K_{C}}}
\sqrt{1+\sqrt{1-3K_{A}K_{C}/\bar{K}_{B}}},
\end{equation}

\noindent provided $1/q^{*}$  is larger
than the lattice or disclination cutoff
size. The thin dotted lines in Figures 5 and 6
show the possible modulated hexatic phase.
Near the critical point,
disclination unbinding and melting
directly to an isotropic fluid is also
possible.

%However, the modulated bond
%angle state requires an intermediate
%hexatic phase.  In a mean field
%approximation for $\varphi$, a bond
%angle modulation instability occurs
%within the hexatic phase when

%\begin{equation}
%g^{2}\left[{1 \over B}+{4\pi-\mu\over 4\pi(\mu+B)-\mu
%B}\right] \geq {\gamma^{2}\over K_{B}}.\label{ggamma}
%\end{equation}

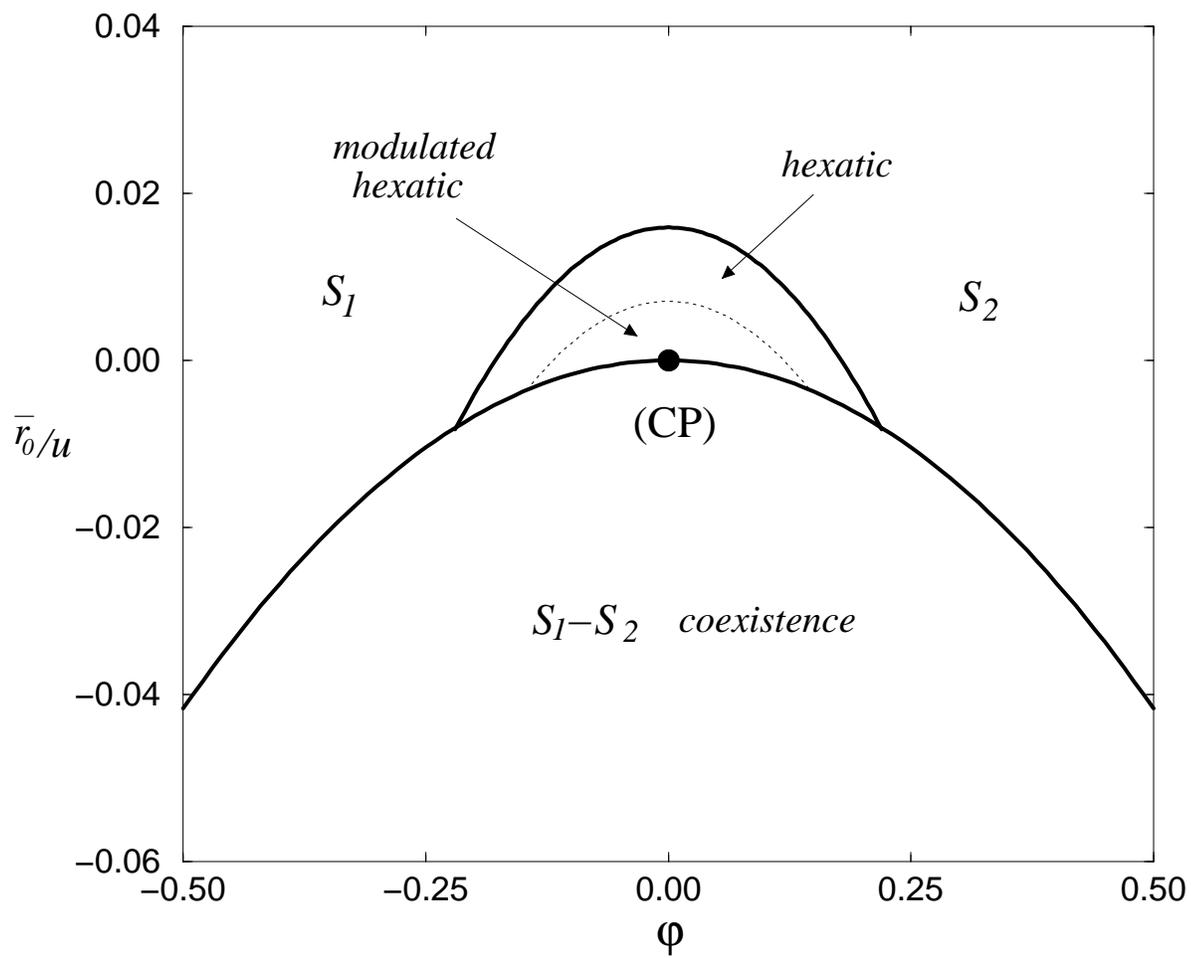
\begin{figure}
\caption{Phase diagram near the critical point of an
isostructural solid-solid transition projected on the
($\bar{r}_{0}/u, \varphi = \rho-\rho_{c}$) plane. For $\mu
=\lambda = 10\pi$, a lobe of hexatic phase is induced
near the critical point. The solid hexatic-solid boundary
corresponds to $g^{2}/u = 2$. The hexatic region becomes
smaller for smaller $g^{2}/u$. A modulated hexatic
phase, evaluated using $v^{2}/K_{B} = 1/45\pi$,
lies within the thin dotted line.
The $S_{1}-S_{2}$ coexistence
line is approximated by mean-field theory.} \label{rhoT}
\end{figure}

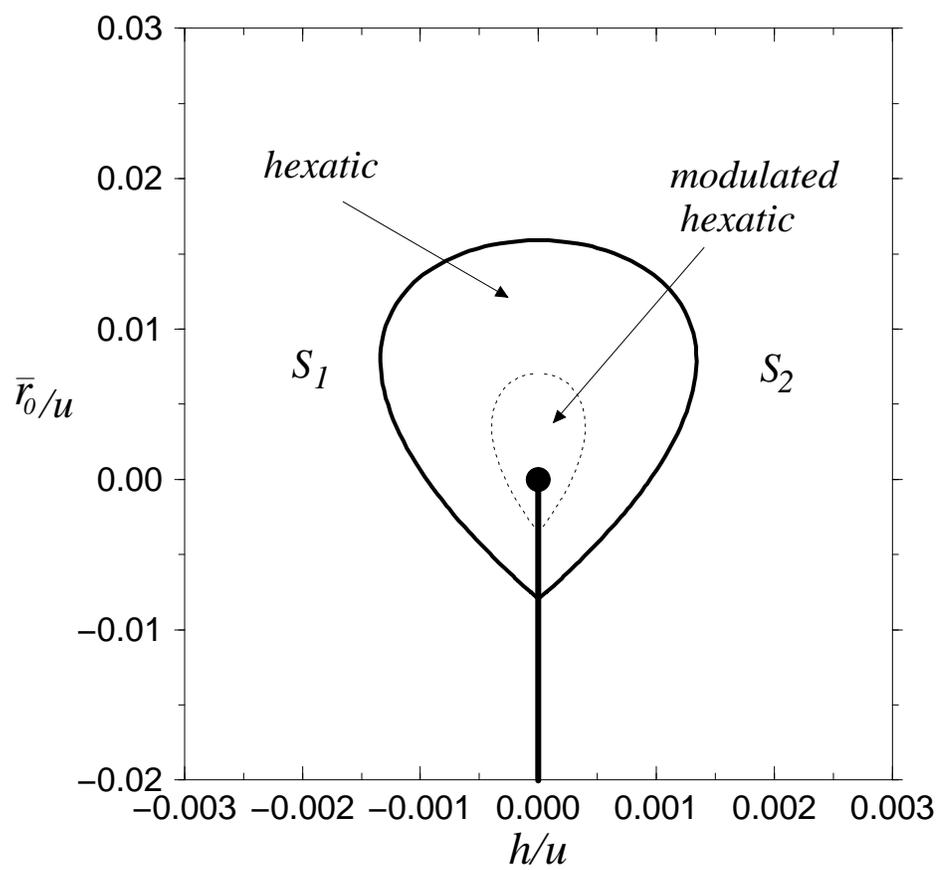
\begin{figure}
\caption{Phase boundaries projected on the
$\bar{r}_{0}/u-h/u$ plane for
$\mu=\lambda = 10\pi$ and
$g^{2}/u = 2.0$.  The smaller lobe corresponding to a
modulated hexatic phase is shown by the thin dotted
line. A line of discontinuous transitions
is represented by the dark line along the $x$-axis
and a renormalized mean-field critical point is
at the origin.}\label{hT1}
\end{figure}

In principle, the solid could melt directly into a
modulated hexatic phase via a first order transition.
Curves delineating possible hexatic-modulated
hexatic and solid-modulated hexatic
phase boundaries are shown in Figures
\ref{rhoT} and \ref{hT1}.

Provided the gradient coefficient
coefficient in (\ref{HV1}) remains positive
the critical point retains non-classical 2D Ising
critical behavior. Coupling to bond angle degrees of
freedom does not induce the long-ranged elastic
interactions that lead to the mean-field critical behavior discussed
in Section III.

\acknowledgements

We are grateful to D. Frenkel for a preprint of
\cite{FRENKEL1} and to B. I. Halperin and C. A. Murray for helpful
discussions.  This work was supported by  the National
Science Foundation, primarily  through the
Harvard Materials Research
Science and Engineering Center via DMR-9400396
and in part through grant No. DMR-9417047.

\appendix

\section{Stability Analysis}

Bounds on the elastic parameters in Eq.
(\ref{Htot}) follow from a stability analysis
in the absence of dislocations. We set $h=0$ for simplicity.
Two cases must
be treated separately.  First, we analyze
stability with respect to the two components
of the displacements $\vec{u}$.  Then,
stability with respect to the three components
of zero wavevector strains is checked.

We first write the Fourier transformed
Hamiltonian after decomposing the
strains into uniform bulk distortions
and finite wavevector modes as in
Reference \cite{LARKIN}.  In the
decomposition (\ref{DECOMP6}), the
uniform symmetric strains $u_{ij}^{(0)}$
must be treated separately because these
require {\it three} degrees of freedom,
in contrast to the two independent modes
when $\vec{q}\neq 0$.  First consider the finite
$\vec{q}$ contribution to the quadratic
part of ${\cal H}$

\begin{equation}
{\cal H}^{(2)}_{\vec{q}\neq 0} = {1 \over 2}\sum_{\vec{q}
\neq 0} u_{i}D_{ij}(\vec{q})u_{j}
+ {1 \over 2}\sum_{\vec{q}\neq 0}(q^{2}+r)\vert
\varphi(\vec{q})\vert^{2} - g\sum_{\vec{q}\neq 0}
iq_{i}u_{i}\varphi(\vec{q}).\label{H2}
\end{equation}

\noindent where the phonon dynamical
matrix is

\begin{equation}
D_{ij}(\vec{q}) = \mu q^{2}\delta_{ij}
 + (\mu+\lambda)q_{i}q_{j}. \label{DM}
\end{equation}

\noindent The eigenvalues of the
quadratic form (\ref{H2}) are given by

\begin{equation}
\begin{array}{c}
\displaystyle \Lambda_{0} = \mu q^{2} \\
\displaystyle \Lambda_{\pm} =
{(2\mu+\lambda)q^{2} + r +
q^{2} \over 2} \pm {1 \over
2}\sqrt{((2\mu+\lambda)q^{2}+r+q^{2})^{2}
-4(2\mu+\lambda)
q^{2}(r+q^{2})+g^{2}q^{2}}.
\end{array}
\end{equation}

\noindent The onset of instability
occurs when an eigenvalue vanishes.
Thus, the system is stable against long
wavelength ($q\rightarrow 0$) modes when
$\mu > 0$, and $r>
g^{2}/(2\mu+\lambda)$.

For uniform distortions, the quadratic
terms in the energy can be written in
terms of the three independent
components of the strain tensor and
$\varphi$

\begin{equation}
{\cal H}^{(2)}_{\vec{q}=0}= \int d^{2}x\, {1 \over
2}(\mu+\lambda)X_{+}^{2}+{\mu \over 2}X_{-}^{2} +
2\mu Y^{2} + g\varphi X_{+} +
\int d^{2}x\, {r \over 2}\varphi^{2} \label{XY}
\end{equation}

\noindent where we have defined

\begin{equation}
\begin{array}{c}
X_{+} = u_{\ell \ell} \\
X_{-} = u_{xx} - u_{yy} \\
Y = u_{xy} = u_{yx}
\end{array}
\end{equation}

\noindent The four eigenvalues in this case are given by

\begin{equation}
\Lambda_{\pm} = {1 \over 4}\left[ B+r \pm
\sqrt{(B-r)^{2}+4g^{2}}\right],\quad \Lambda_{1} =
2\mu,\quad \Lambda_{2} =  {\mu\over 2}
\end{equation}

\noindent where $B\equiv \mu+\lambda$.
Stability requires $\mu > 0$, and

\begin{equation} r >
g^{2}/(\mu+\lambda).\label{STABILITY}
\end{equation}

\noindent Thus, as in a pure elastic
theory, the uniform strains impose a
more stringent criteria on stability
than do the finite wavevector modes.
Therefore, as temperature is lowered ($r=a_{2}(T-T_{c})$), the
uniform mode $\varphi_{0}$ becomes
unstable first.

\end{document}